\documentclass{iau}
\usepackage{graphicx}

\newcommand{\Rsun}{R_{\odot}}

\title[Fragmentation and Core Collapse] 
{Molecular Cloud Fragmentation\\ and Core Collapse}

\author[S. Basu]   
{Shantanu Basu} 

\affiliation{Department of Physics and Astronomy, The University of Western Ontario, \\ 
London, Ontario N6A 3K7, Canada\\ email: {\tt basu@uwo.ca}}

\pubyear{2015}
\volume{315}  
\pagerange{XXX--XXX}
\jname{From interstellar clouds to star-forming galaxies: universal processes?}
\editors{P. Jablonka, P. Andr\'e, \& F.. van der Tak, eds.}
\begin{document}

\maketitle

\begin{abstract}
I review some steps in the conversion of molecular cloud gas into stars and planets,
with an emphasis in this presentation on the early stage molecular cloud fragmentation
that leads to elongated filaments/ribbons. 
Magnetic fields can play a crucial role in all stages and need to be invoked 
particularly for early stage fragmentation as well as 
in late core collapse where it may control disk formation. I also review some 
elements of hydrodynamic modeling of disk evolution.

\keywords{ISM: clouds, ISM: magnetic fields, ISM: molecules, MHD, stars: formation}
\end{abstract}

\firstsection 
\section{Introduction}

The conversion of molecular gas into stars and planets can for simplicity
be conceptually divided into four stages: (1) the fragmentation of clouds into
large scale structures (filaments/ribbons); (2) the formation of dense cores within
these large structures; (3) core collapse to form a hydrostatic protostar; (4) disk
formation and evolution leading to outflows, multiplicity, planets, etc. In this review 
I emphasize step (1), given recent developments in the literature, and comment 
briefly on the others.


\section{Cloud Fragmentation and Filaments}

\begin{figure}
\centering
\includegraphics[width=90mm]{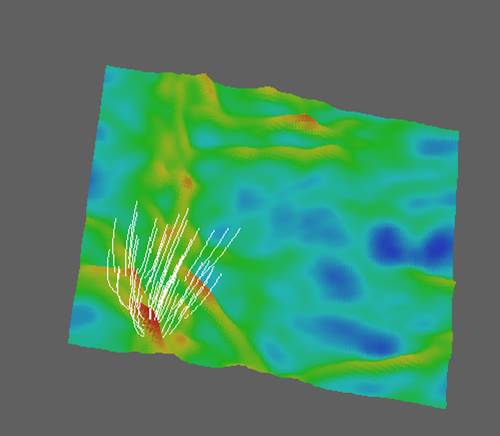}
\caption{An image of column density in a thin sheet perturbed with turbulent initial 
conditions. There is a large scale magnetic field initially perpendicular to the sheet,
and is illustrated with white lines in the maximally compressed region only. Turbulence
and magnetic fields lead to elongated structures (ribbons) with magnetic field largely
perpendicular to the long axis of the ribbon. From a simulation by
\cite[Basu et al. (2009)]{bas09}. \label{scenario}}
\end{figure}

Molecular clouds are assembled from H {\small I} clouds in the interstellar
medium that have a mass-to-flux ratio $M/\Phi$ that is significantly less than the
critical value required for gravitational fragmentation and collapse
(\cite[Heiles \& Troland 2005]{hei05}). If molecular clouds are assembled from these
subcritical components, then the observed star formation in the clouds leads
to two possibilities: (1) the molecular clouds are entirely supercritical, having been
assembled from a sufficient amount of subcritical material, by motions along Galactic
magnetic field lines; (2) the molecular clouds are largely subcritical but that the 
breakdown of flux freezing due to ambipolar diffusion allows subregions to become
subcritical and produce dense cores. The first scenario places challenging demands on 
the accumulation length of matter in molecular clouds, and also on the speed of
streaming motions along the magnetic field (\cite[Mestel 1999]{mes99}). The second
scenario has has the drawback that the ambipolar diffusion
timescale is at least $~10^7$ yr, or even much longer, in diffuse molecular gas. 
However, the addition of turbulent initial conditions leads to the formation of
dense structures on a much shorter turbulent crossing time. Within the compressed gas,
rapid ambipolar diffusion
can take place due to the elevated density and magnetic field gradient. This scenario
has been explored in many studies over the last decade 
(\cite[Nakamura \& Li 2005]{nak05}; \cite[Elmegreen 2007]{elm07};
\cite[Kudoh \& Basu 2008]{kudo8}; \cite[Nakamura \& Li 2008]{nak08};
\cite[Basu et al. 2009]{bas09}; \cite[Kudoh \& Basu 2011]{kud11})
and also provides a natural way to understand filamentary structures in clouds.
By filaments we do not mean objects with cylindrical symmetry, but rather more
ribbon-like structures, with flattening along the ambient magnetic field direction, and 
compression by turbulence in primarily one direction perpendicular to the magnetic
field. The overall scenario is illustrated in Figure~\ref{scenario}, taken from a simulation
of \cite[Basu et al. (2009)]{bas09}. The filament formation is thus quite distinct from the cosmological 
filament formation simulations (e.g., \cite[Springel et al. 2005]{spr05}). 
in which the gravitational field of cold dark matter is able to collapse into filamentary
structures on a local dynamical time. 
The magnetic turbulent picture allows the formation of filaments but not necessarily 
in a volume filling network as in cosmological simulations, since large parts of clouds 
may remain in a subcritical common envelope with little to no star formation. Models of
magnetohydrodynamic (MHD) wave propagation in inhomogeneous clouds 
(\cite[Kudoh \& Basu 2003]{kud03}; \cite[Kudoh \& Basu 2006]{kud06}) also show that the 
velocity dispersion is essentially Alfv\'enic and that low density regions with high 
local Alfv\'en speed support large amplitude velocity fluctuations, in agreement with 
the observed linewidth-size relations of molecular clouds
(\cite[Solomon et al. 1987]{sol87}). Magnetic fields can also 
lower the dissipation rate of turbulence in some cases, although not necessarily 
dramatically (\cite[Kim \& Basu 2013]{kim13}), and allow more efficient distribution of outflow power 
across the cloud (\cite[Wang et al. 2010]{wan10}).

Recent observations by the 
{\it Herschel Space Observatory} have established that the dense cores and young
stellar objects are often threaded along networks of elongated structures seen 
in projection (\cite[Andr\'e et al. 2010]{and10}; 
\cite[Men'schikov et al. 2010]{men10}).
\cite[Arzoumanian et al. 2011]{arz11}).
When magnetic field morphology can be detected through polarized dust emission,
the magnetic field is generally aligned perpendicular to the long axis of the
filaments (\cite[Palmeirim et al. 2013]{pal13}). These features are at least qualitatively
consistent with the magnetic turbulent scenario described above. Another interesting
result is that the measurement of average FWHM widths across filaments in three 
different clouds shows a cluster of values around $\sim 0.1$ pc, even though the 
Jeans length calculated from the peak column densities are varying by about
two orders of magnitude (\cite[Arzoumanian et al. 2011]{arz11}). 

It seems that the
filament widths require an explanation that goes beyond the simple Jeans length
prescription. They are likely to be quasi-equilibrium objects based on the amount
of fragmentation along their lengths to form dense cores. Recall that for a 
non-magnetized isothermal cylinder, there is a critical line mass $2 c_s^2/G$ (where 
$c_s$ is the isothermal sound speed) below which radial 
equilibrium is possible but fragmentation occurs along the axis of the cylinder.
However if the line mass exceeds this critical value, then indefinite collapse to a
spindle occurs before any fragmentation can take place 
(\cite[Inutsuka \& Miyama 1997]{inu97}). Modified values of critical line mass, actually
critical line mass to line flux ratio, for magnetized cylinders with a lateral
magnetic field have recently been obtained by \cite[Tomisaka (2014)]{tom14} and
\cite[Hanawa \& Tomisaka (2015)]{han15}. Given the fragmentation along their
length, it is likely that the 
line mass to line flux ratio of observed ribbons is subcritical.

\cite[Hennebelle \& Andr\'e (2013)]{hen13} developed an analytic
model of filament width in which the energy input from accretion onto a filament 
is balanced by energy 
dissipation by ion-neutral friction, and there is mechanical equilibrium between
gravity and internal turbulence. They obtained a mean filament width of $\sim 0.1$ pc
for a range of column densities. An alternate model of filament formation was
explored by \cite[Kudoh \& Basu (2014)]{kud14} in which a filament (really ribbon) is 
created by turbulent compression of subcritical gas and halts contraction when 
magnetic pressure balances ram pressure. However, their semi-analytic model was 
used to estimate time scales for core formation and not to estimate filament widths.


\section{Cores to Star-Disk Systems}

\begin{figure}
\centering
\includegraphics[width=90mm]{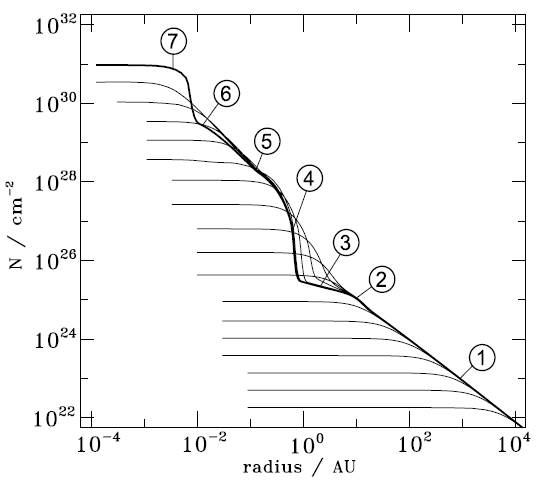}
\caption{
	Column number density ($N$) profile versus radius from 
	\cite[Dapp et al. (2012)]{dap12}. The thin lines are profiles in
ascending order of time. Several features are identifiable via their associated breaks in the profile
(1) Prestellar infall profile with $N \propto r^{-1}$. (2) Magnetic wall at
$\approx 10~\mathrm{AU}$, where the bunched-up field lines
decelerate material before it continues the infall. (3) Expansion wave profile with
$N \propto r^{-1/2}$ outside the first core.
(4) First core at $1~\mathrm{AU}$. (5) Infall profile onto the second core with $N \propto r^{-1}$.
After the first core has reached $\approx 1,000~\mathrm{K}$, it starts to collapse, as
H$_{2}$ is dissociated. (6) Expansion wave profile with
$N \propto r^{-1/2}$ outside the second core.
(7) Second core at $\approx 1~\Rsun$.
\label{profiles}}
\end{figure}

\begin{figure}
\centering
\includegraphics[width=75mm]{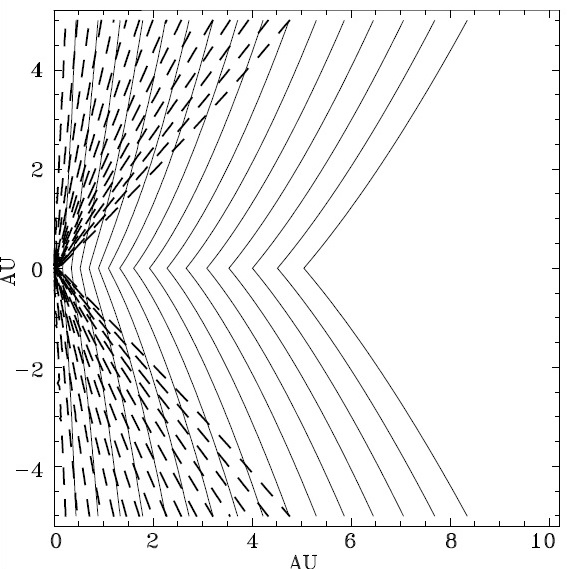}
\caption{
	Magnetic field lines on $10~\mathrm{AU}$ scales from \cite[Dapp et al. (2012)]{dap12}.
The dashed lines represent the flux-freezing model, while the solid lines show \textit{the same} field lines for
the model including non-ideal MHD effects for a grain size $a_{\mathrm{gr}}=0.038~\mathrm{\mu m}$. 
In both cases, the second core has just formed and is on the left axis midplane. 
The field lines straighten out significantly on small scales in the non-ideal MHD model compared to the 
flux-frozen model.
\label{flines}}
\end{figure}

Once a dense core is formed by fragmentation of a larger clump
or filament, it will begin runaway collapse as long as its 
mass-to-flux ratio is supercritical. The hydrodynamic collapse
leads to a well known set of outcomes: development of an $r^{-2}$
density profile and a shrinking central flat density peak that
eventually forms the first hydrostatic core, later collapsing
to the second core of stellar dimensions (e.g., \cite[Larson 1969]{lar69}).
The addition of rotation and/or magnetic fields leads to a flattened
collapsing core, especially at high densities. Once a central 
protostar is formed at the pivotal moment $t=0$, several interesting 
things happen. An expansion wave moves outward from the center 
(\cite[Shu 1977]{shu77}), changing the character of flow variables within 
it. When rotation is added, a centrifugal disk is able to form, 
however an additional flux-frozen magnetic field leads to catastrophic
magnetic braking that shuts off disk formation (\cite[Allen et al. 2003]{all03};
\cite[Galli et al. 2006]{gal06}). 

A straightforward solution to catastrophic magnetic braking is the presence of 
non-ideal MHD effects, notably ambipolar diffusion and Ohmic dissipation.
These will weaken the magnetic field strength and reduce the effectiveness
of magnetic braking. In \cite[Dapp \& Basu (2010)]{dap10} and 
\cite[Dapp et al. (2012)]{dap12} we used the thin-disk approximation to follow
collapse to the formation of a second stellar core while including non-ideal
MHD effects. Both studies show that non-ideal MHD effects allow the formation
of a disk immediately after the pivotal moment $t=0$. In \cite[Dapp et al. (2012)]{dap12}
we used a chemical network model to calculate the ionization fraction
and coefficients of ambipolar diffusion and Ohmic dissipation. Figure~\ref{profiles}
shows a series of column density profiles at various times, with the last one
at the moment immediately following the formation of a second core of 
stellar dimensions. Each distinct region of the column density profile is captured,
including the MHD accretion shock created by outward moving magnetic flux 
(\cite[Li \& McKee 1996]{li96}; \cite[Contopoulos et al. 1998]{con98}).
Figure~\ref{flines} shows the magnetic field lines in cases with 
ideal and non-ideal MHD. The former case has extreme flared field lines 
and no disk is able to form due to catastrophic magnetic braking.
The non-ideal MHD model yields the development of a centrifugal disk immediately after the
formation of the second core, however the loss of some angular momentum due to 
magnetic braking is expected to
keep disks at relatively small sizes $\lesssim 10$ AU in the early Class 0 phase
($\lesssim 4 \times 10^4$ yr).

\section{Disks and Multiplicity}

\begin{figure}
\centering
\includegraphics[width=90mm]{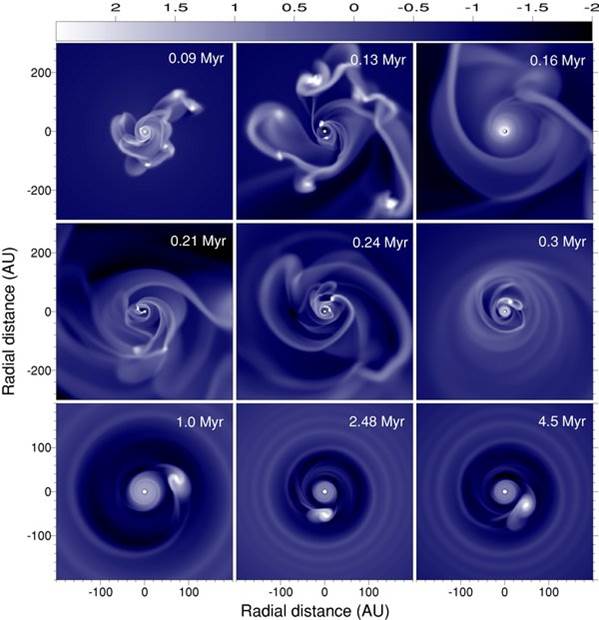}
\caption{
Gas surface density maps (g~cm$^{-2}$, log units) at six times after
      the formation of the central star (bright circle in the coordinate center), from
      \cite[Vorobyov \& Basu (2010a)]{vor10a}.
      Note the zoom-in as the time increases. The top two rows contain images of size 600 AU on
      each side, while the bottom row contains images of size 400 AU on each side.
\label{disk}}
\end{figure}

As shown in the previous section, the collapse of a
prestellar core leads to a very focused
collapse with a power-law density profile. As a result,
the probability of fragmentation during
the runaway collapse phase is low.
However, a centrifugally supported disk can form after a central protostar has
formed. The existence of a central point mass potential allows
a centrifugal barrier to be encountered by an infalling rotating fluid element
that conserved its angular momentum (see discussion in 
\cite[Basu \& Mouschovias 1995]{bas95}). 
A protostellar disk is a quasi-equilibrium structure and evolves on a time scale much longer than
its dynamical time.
Therefore, it has time for fragmentation if conditions
are suitable, e.g., the criteria for gravitational instability and fragmentation are
satisfied (\cite[Toomre 1981]{too81}). 
Disk fragmentation is the most likely mechanism for the 
formation of multiple stellar systems and the formation of brown dwarfs and giant planets at
wide orbits of several tens of AU. 
We show here the results of hydrodynamic numerical simulations of 
\cite[Vorobyov \& Basu (2010a)]{vor10a}, who modeled the formation
of the disk from the collapse of a prestellar core, as well as its subsequent evolution on
Myr time scales. Figure~\ref{disk} shows images of the disk column density at a series of times 
after formation of the central protostar. During the first $10^5$ yr, there is 
vigorous episodic gravitational
instability in the disk, driven by mass accretion from the parent core. Clumps form in spiral
arms through nonlinear gravitational instability, and are eventually driven in to the center
through gravitational interaction with trailing spiral arms. In this model, one clump is able
to survive to late times and eventually carves out a gap in the disk and settles into a stable
orbit by 4.5 Myr. The clump is essentially a first hydrostatic core and is 
the progenitor of a low mass companion star with an orbit radius $\approx 52$ AU. This mechanism
can also produce wide orbit proto-brown-dwarfs and giant protoplanets. 
Details of the numerical model and the overall scenario of the migrating embryo model can
be found in 
\cite[Vorobyov \& Basu (2006)]{vor06}, \cite[Vorobyov \& Basu (2010b)]{vor10b}, and
\cite[Vorobyov \& Basu (2015)]{vor15}.

\end{document}